\documentclass{article}
\usepackage{emulateapj,psfig}

\def\simlt{\mathrel{\spose{\lower 3pt\hbox{$\mathchar"218$}}
     \raise 2.0pt\hbox{$\mathchar"13C$}}}
\def\simgt{\mathrel{\spose{\lower 3pt\hbox{$\mathchar"218$}}
'     \raise 2.0pt\hbox{$\mathchar"13E$}}}
\def\gsim{ \lower .75ex \hbox{$\sim$} \llap{\raise .27ex \hbox{$>$}} }
\def\lsim{ \lower .75ex \hbox{$\sim$} \llap{\raise .27ex \hbox{$<$}} }

\long\def\***#1{{\scshape ***#1***}}
\def\arraystretch{1.25}

\makeatletter

\newenvironment{inlinefigure}{%
\def\@captype{figure}%
\noindent\begin{minipage}{0.999\linewidth}\begin{center}}
{\end{center}\end{minipage}\smallskip}
\makeatother

\begin{document}
\lefthead{Quilis and Moore}
\righthead{Where are the HVCs?}

\title{Where are the high velocity clouds?}

\author{Vicent Quilis and Ben Moore}

\affil{Physics Department, University of Durham, Durham DH1 3LE, UK}

\begin{abstract}
Recent observations of high velocity clouds (HVCs) have revealed compression
fronts and tail shaped features of HI suggesting that they are interacting with an
external medium.  We perform 3-D hydro-dynamical simulations 
of HVCs moving through a diffuse hot gaseous component, investigating 
the behaviour
of both extra-galactic dark matter dominated HVCs and nearby
pure gas clouds that may be accreting onto the Galactic disk 
via a Galactic fountain. Both scenarios can give rise to similar
features as observed if the external medium has a density $>10^{-4}$ cm$^{-3}$.
Observations suggest that this may be too high for 
a hot ionised halo or intergalactic gas and supports the Galactic
fountain origin, a model that can also account for the high fraction of HVCs
with tails and asymmetrical morphologies.
\end{abstract}


\section{Introduction}

The nature of the high-velocity clouds (HVCs) seen in 21 cm HI emission remains
controversial since their discovery over three decades ago (Muller, Oort \& 
Raimond
1963).  Most fundamentally, the distances, hence mass and environment of these
objects remain poorly constrained, largely due to the difficulty of finding
background sources (against which the clouds would appear in absorption) at
known, comparable distances. It is possible that the
HVCs may not have a single common origin, but rather 
they are a mix of relatively nearby ($|z| < 10\,{\rm kpc}$), ``Galactic
fountain'' clouds, of material stripped from the Magellanic Clouds and other
satellites of the Milky Way, and perhaps of extra-galactic objects (Blitz et
al. 1999; Braun \& Burton 1999, 2000).  This latter possibility is most 
intriguing
since the HVCs would provide a link to the numerous dark matter substructure
halos expected in the hierarchical cold dark matter model 
(Moore et al. 1999, Klypin et al. 1999).

Previous studies have used the morphological, two phase core-envelope structure
of the HVCs to infer approximate distances (Wolfire et al. 1995).  
Bland-Hawthorn \& Maloney (1999) have proposed the use of 
sensitive $H_{\alpha}$
emission measurements, together with a model for the ionizing photon flux as a
function of position within the Galactic halo to infer their distances. A
handful of HVCs have upper distance limits that are $\lsim 10$ kpc
due to the presence of background stars (c.f. Gibson et al. 2000).

Br\"uns et al. (2000, 2001) found that $\approx 20\%$ of HVCs
exhibit position-velocity gradients, termed
\emph{head-tail} HVCs by analogy with the structures of comets. 
A likely cause for their elongated morphology is that they are interacting
with a more diffuse, ionized ambient wind. Further evidence for such cloud-halo
interactions are evident from the compressed leading fronts of many HVCs.
In this letter we use high resolution hydro-dynamical simulations to examine 
the
morphologies of HVCs moving through a diffuse medium. This allows us to infer
the efficiency of the stripping process and will place limits on the possible
environment of the HVCs.  
A detailed study of the interaction between clouds and a diffuse medium
can be found in Murray et al. (1993).

\section{The structure of the HVCs}

The typical angular size of HVCs are $\theta\sim 1^{\circ}$ and in order to
be included in existing catalogues (eg, Br\"uns et al. 2000) their
HI column density, $N_{HI}$, should be $\gsim 10^{19}$ cm$^{-2}$.
For a cloud at distance $D$, we can estimate its
radius, $R$, mean spatial HI density,
$n_c$, and mass, $M_c$, as follows.
$R=\theta D \sim 2 {\rm kpc} (\theta/1^{\circ})(D / 100{\rm kpc})
$
and $n_c={N_{HI}/2R}$, i.e.
$$
n_c\simeq 10^{-3}\, {\rm cm}^{-3} \left(N_{HI}\over
10^{19}\, {\rm cm}^{-2}\right) \left(\theta \over 1^{\circ}\right)^{-1}
\left(D\over 100\, {\rm kpc}\right)^{-1}
$$
The cloud mass $M_c \simeq f_{HI}^{-1} N_{HI} \pi R^2 m_{HI}$, can be written;
$$
M_c\simeq 2\times10^5 f_{HI}^{-1}
M_{\odot}
\left( N_{HI} \over 10^{19}\, {\rm cm}^{-2}\right) \left(\theta \over
1^{\circ}\right)^2 \left(D \over 100\, {\rm kpc}\right)^2
$$
\noindent
where $f_{HI}$ is the neutral hydrogen fraction.
A cloud is gravitationally bound when it satisfies the virial relation
$R \sigma_c^2 \simeq GM_c$. With our model assumptions
and $\sigma_c \sim 10$ km/s, this
translates to
$$
10^{12} \, \frac{{\rm cm}^2}{s^2} = 2\times 10^{10} \frac{{\rm cm}^2}{f_{HI}\, 
s^2}
\left(N_{HI}\over 10^{19}\, {\rm cm}^{-2}\right)
\left(\theta \over 1^{\circ}\right) \left(D \over 100\, {\rm kpc}\right)
$$.

The general conclusion is that the nearby HVCs at distances
$D=1-10$ kpc, have masses $M_c=10-100M_\odot$ and must be in
pressure equilibrium with the warm ionised medium located
within and above the galactic disk. HVCs at several hundred kpc may
also be pressure confined by some external medium,
or they may be gravitationally bound by a dark matter halo. 
For both type of clouds we assume that they consist 
mainly of warm gas with temperature $T_c \sim 10^4\, K$, velocity dispersion 
$\sigma_c \sim 10$ km/s, and density $n_c$.
The dark matter dominated HVCs are constructed by adopting a
dark matter density profile $\rho(r)={A}/{(r_c^2+r^2)}$ 
where $A={v_{cir}^2}/{4\pi G}$, $v_{cir}=15$ km/s and $r_c=1$ kpc. 
The HI lies primarily within the constant density core of the
dark matter halo and we adjust its temperature profile to maintain 
hydro-static equilibrium.
A typical cloud may be at a distance of 300 kpc, have a radius of 
$\approx 1$ kpc and a total gas mass of $1.6\times 10^7M_\odot$. 
This leads to a central column density of $10^{20}$ cm$^{-2}$ 
which is higher than the average quoted by Wakker and van Woerden (1991)
but similar to those observed by Br\"uns et al. (2001).

The diffuse halo material (number density $n_w$) is taken to be at a 
temperature $T_h \sim 10^6\, K$, comparable to the
virial temperature for a circular velocity of 200 km/s.
The wind velocity $v_w$ should be of this same order; this is
confirmed by the observed distribution of HVC galactocentric velocities.
Clouds ejected from the disk in a Galactic fountain
would fall back to the disk with a velocity
$\approx \sqrt{GM_{disk}/z} \sim 200$km/s, 
where $M_{disk}\approx 5 \times 10^{10}M_\odot$ and vertical height
 $z\approx 10$ kpc.

\section{The density of diffuse gas in galactic halos}

The properties of the diffuse outer halo of the Milky Way is poorly known. Such a
halo is expected to exist as left over material that has been unable to cool 
since the
formation of the Milky Way.  Observations so far provide only upper
limits on its mean density. 
Constraints come from the measured soft X-ray
background (Snowden et al. 1997, Benson et al. 2000), from the dispersion 
measures of pulsars in globular 
clusters and in the Magellanic Clouds, from the 
kinematics (Moore \& Davis 1994) or
lifetime (Murali 2000) of the Magellanic Stream.
All of these constraints place the density of the ionised Galactic halo gas
$n_w(50{\rm kpc})\lsim 5\times10^{-5} {\rm cm}^{-3}$.  
Blitz \& Robishaw (2000) derive a halo gas density that is a factor of 2 
lower than this at a 
distance of 200 kpc by applying ram pressure stripping to the halo dwarf 
spheroidals.
Deep ROSAT observations of nearby disk galaxies
provide complimentary upper limits on a very extended component, but they give
positive detections of ionised gas close to the disks that has a density of
$n_w(10{\rm kpc})\approx 10^{-3} {\rm cm}^{-3}$ 
(Bregman \& Houck 1997, Ehle et al. 1998). We caution that 
direct observational limits of the
diffuse gas density at 200-500 kpc do not exist.

The fact that the HVCs are moving at 150-300 km/s through this medium 
provides an
upper limit to its density by considering their terminal velocity 
(c.f. Benjamin \& Danly 1997).
The equation of motion of a pure 
gas cloud of mass $m_c$, area $A$ and 
velocity $v_c$ through a diffuse medium of density $\rho_w$ is
$m dv_c/dt = \rho_w(r) v_c^2 A - m_c g(r)$.
Assuming that the gas is at terminal velocity will give an 
upper limit to $\rho_w(r)$. In this case $dv_c/dt=0$ and 
$v_t=\sqrt{{N_{HI}.(GM/r^2)}/{\rho_w(r)}}$, therefore a pure gas
cloud with column density $N_{HI}=10^{20}$ cm$^{-2}$, at distances of
10 kpc and 100 kpc moving at $v_t=300$ km/s constrains the local IGM 
$n_w(10{\rm kpc})<3\times 10^{-3}$ cm$^{-3}$
and $n_w(100{\rm kpc})<3\times 10^{-4}$ cm$^{-3}$ (for pure gas clouds).

The amount of gas stripped from a dark matter dominated HVC can be
estimated using the analytic approach described in 
Gunn \& Gott (1972). The gaseous component will be stripped 
when the ram pressure of the environment is larger than the gravitational 
restoration force, 
$\rho_w v^2_w= ({\partial\phi_{_{DM}}}/{\partial r}) \sigma$ where $\sigma$ is 
the surface density. This is plotted in Figure 1 for our fiducial
dark matter dominated HVC and predicts that it will survive intact 
for halo densities
$n_w<10^{-5}$ cm$^{-3}$ but would be completely stripped if $n_w$ is an 
order of
magnitude larger than this. 
Figure 1 also shows the results for gas clouds embedded within a
cuspy density profile with $\rho(r) \propto r^{-1}$. The steeper potential 
can retain
gas more easily and may explain why tiny dwarf galaxies such as Fornax can hold on to gas
and form new stars at its center.

\begin{inlinefigure}
\centerline{\includegraphics[width=1.1\linewidth]{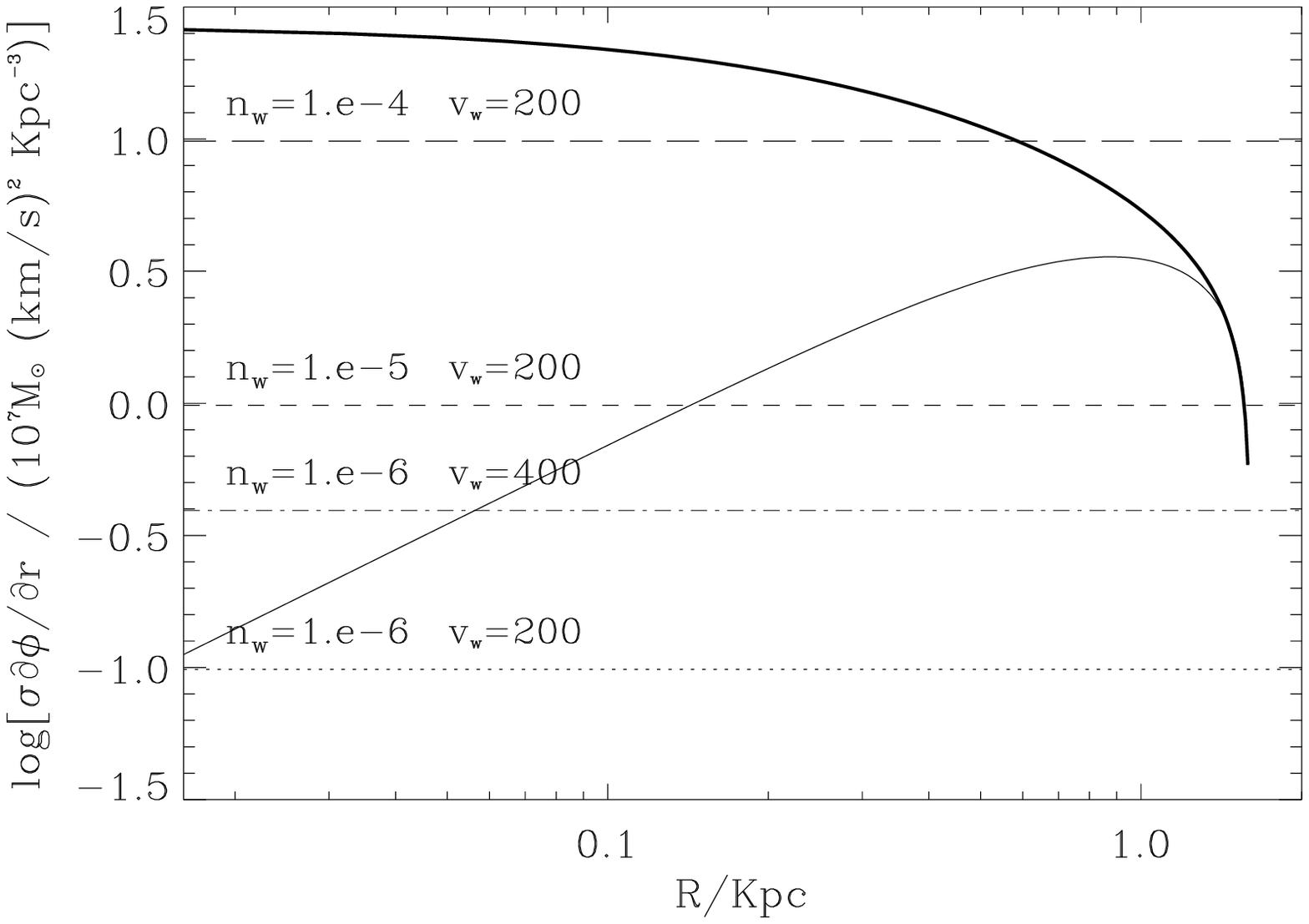}}
\caption{
The restoration force at different radii within a dark matter dominated HVC.
The thin solid curve is for dark matter halos with constant density core
with radius 1 kpc and the thick curve
shows the behaviour of a density profile with a 
central cusp $\rho(r)\propto r^{-1}$.
The far right location where the horizontal lines 
intersect the restoration force indicates the
stripping radius beyond which all the gas will be removed.
}
\label{fig1}
\end{inlinefigure}

\section{Results}

\begin{figure*}
\centerline{\epsfysize=3.5truein \epsfbox{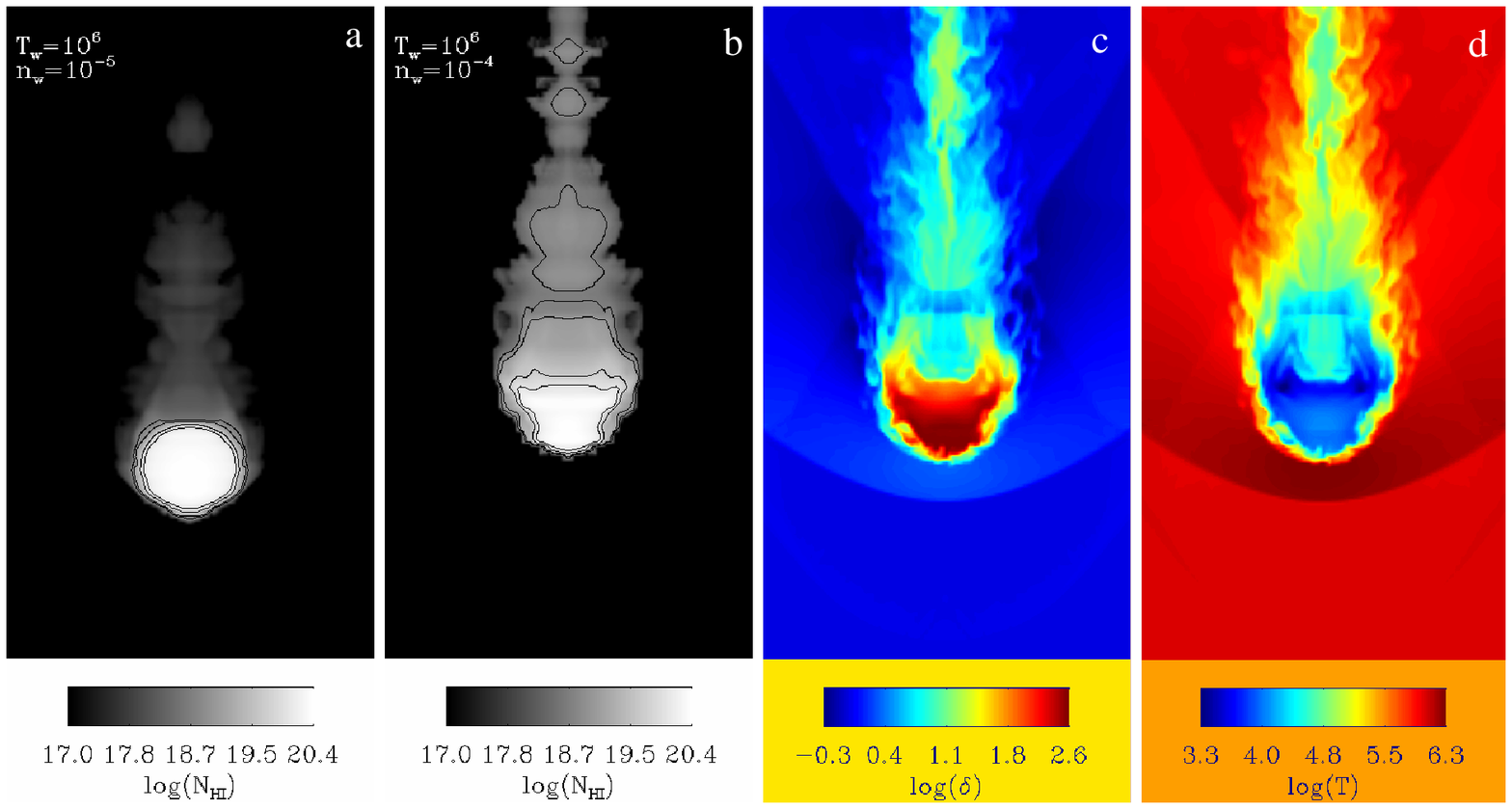}}
\caption{The column density for the dark matter dominated
HVC after 300 Myrs of evolution in a wind of velocity 200 km/s, 
(a) density $10^{-5}$ cm$^{-3}$ and (b) density $10^{-4}$ cm$^{-3}$.
The contours correspond to column densities   
$5, 10.0, 50.0, 100.0 \times 10^{18}$cm$^{-2}$.
Panels (c) and (d) show the physical density ($\delta=\rho/\rho_w$)
and temperature within a central slice through the HVC in panel (b).
} 
\label{fig2}
\end{figure*}

The simulations have been tackled as pure hydro-dynamical processes. 
The HI is described by a
perfect fluid with equation of state 
$p=(\gamma-1)\rho\epsilon$ where $\rho$ is the density, $\epsilon$ the specific 
internal energy, and the adiabatic exponent $\gamma=5/3$. 
The pure gas HVCs are constructed to be in pressure equilibrium with
the external hot medium, whereas the dark matter dominated clouds have
an additional potential field that maintains their 
dynamical equilibrium. In order to integrate 
the hydro-dynamic equations describing the evolution of 
HVCs moving through an ionised medium  
we have used the high-resolution shock-capturing 
method described in Quilis, Ib\'a\~nez and S\'aez (1996).
This code was specially designed for an accurate 
treatment of fluid dynamical processes, and is
extremely good at dealing
with shocks, strong discontinuities, 
turbulent regions and low density regimes. A specially 
adapted version for the HVC simulations has been developed.
We employ a fixed grid with ratio of sides 2:1:1 and a wind blowing
down the long axis. The timestep is set by the minimum Courant condition
in the simulation volume which typically sets several hundred steps
per crossing time.

We ran a grid of simulations, varying the density and temperature of the 
external
medium from $10^{-3}$ to $10^{-6}$ cm$^{-3}$ and $10^6-10^4$K respectively,
and cloud velocities between 170 km/s and 400 km/s in order to be conservative.
Most of our simulations were run without allowing radiative cooling, although
a test run with cooling switched on showed no difference in the response of the
HVC to the wind. Our model HVCs are simplistic since they have no internal 
structure, however we do not expect that this would greatly influence 
our conclusions.

\subsection{I. Dark matter dominated clouds}

Figure 2 shows snapshots of the dark matter dominated HVC 
after $3\times 10^8$ years
which is approximately 20 crossing times ($t_{cr}\sim {2R}/{v_w}$)
at a wind velocity $v_w=200$ km/s.
The physical size of the box is 12 kpc$\times$12 kpc$\times$24 kpc, 
and the number of cells used $128\times128\times256$, giving
a nominal numerical resolution of 90 pc.
We ran several test simulations with $256\times256\times512$ cells and 
found no difference with the lower resolutions runs.  
The grey scale shows the projected surface density of neutral gas and we
plot column density contours of cold HI. We define a tail as visible if it
has column density higher than $10^{19} {\rm cm}^{-2}$.
For wind densities $\leq 2\times 10^{-5} {\rm cm}^{-3}$ we find no 
visible tails, 
although very weak features with $N_{HI}\sim 10^{18}$ cm$^{-2}$ are present.

Panel (b) of Figure 2 shows the same dark matter dominated HVC moving
though a medium an order of magnitude denser. Although this may be 
unrealistic for
a Galactic halo component, it does give rise to visible tails of gas.
All of the neutral gas has been pushed away from the dark matter
potential, as expected from the application of the Gunn and Gott stripping
criteria. This material will eventually be completely stripped away and 
evaporate
within the hot component. 

In all cases that we simulate we find a similar behaviour.
A weak bow shock is visible in front of the cloud that penetrates the IGM
and increases its local temperature. The neutral gas at
the front edge of the HVC is compressed and heated, whilst 
a region of low pressure is created immediately behind the cloud.
Panels (c) and (d) of Figure 2 shows the rich structure and detail that
can be captured with our code.
If the wind is not strong enough to cause ram-pressure stripping, 
viscous/turbulent stripping will continue to remove gas, however, the timescale
for this process is fairly long, hence the tails can be quite weak.

\subsection{II: Pure gas clouds}
 
Figure 3 shows the evolution of the pure gas clouds.  In each case, the 
physical
size of the box is 180 pc$\times$180 pc$\times$ 360 pc the number of grid cells
used is $128\times128\times256$, giving a numerical resolution of 3 pc.
The evolution and features are similar as the dark matter dominated clouds;
when the wind is dense enough then visible tails of stripped HI are created.

If the initial column density is too low, the pure gas clouds are
compressed at the leading edges and the velocity of the entire cloud decreases
as it moves towards its terminal velocity. The main difference between
the HVCs in Figure 3(a) and 3(d) are their initial gas densities,
which is an order of magnitude higher in Figure 3(d). The IGM density 
and velocities are the same, but the denser HVC shows visible tails 
whereas the tails in Figure 3(a) are below $10^{19}$ cm$^{-2}$ and would 
not be observable.

The mass ratio of gas in the head and tails is typically around 10:1.
This result appears very robust regardless of the parameters 
used in the simulation and is lower than the 3:1 ratio 
quoted by Br\"uns et al. (2000).
The characteristic lifetimes of the pure gas clouds
are quite short and once they enter a diffuse environment that is sufficiently
dense to create visible tails, they are rapidly destroyed on a timescale of
$\approx 10^7$ Myrs. Therefore the high fraction of observed clouds with tails
suggests that we are observing them soon after they are impacting into a 
region of high density gas. The tails trail directly behind the motion of 
the clouds
therefore we might expect an alignment effect as the clouds fall directly 
onto the disk.

\begin{figure*}
\centerline{\epsfysize=3.5truein \epsfbox{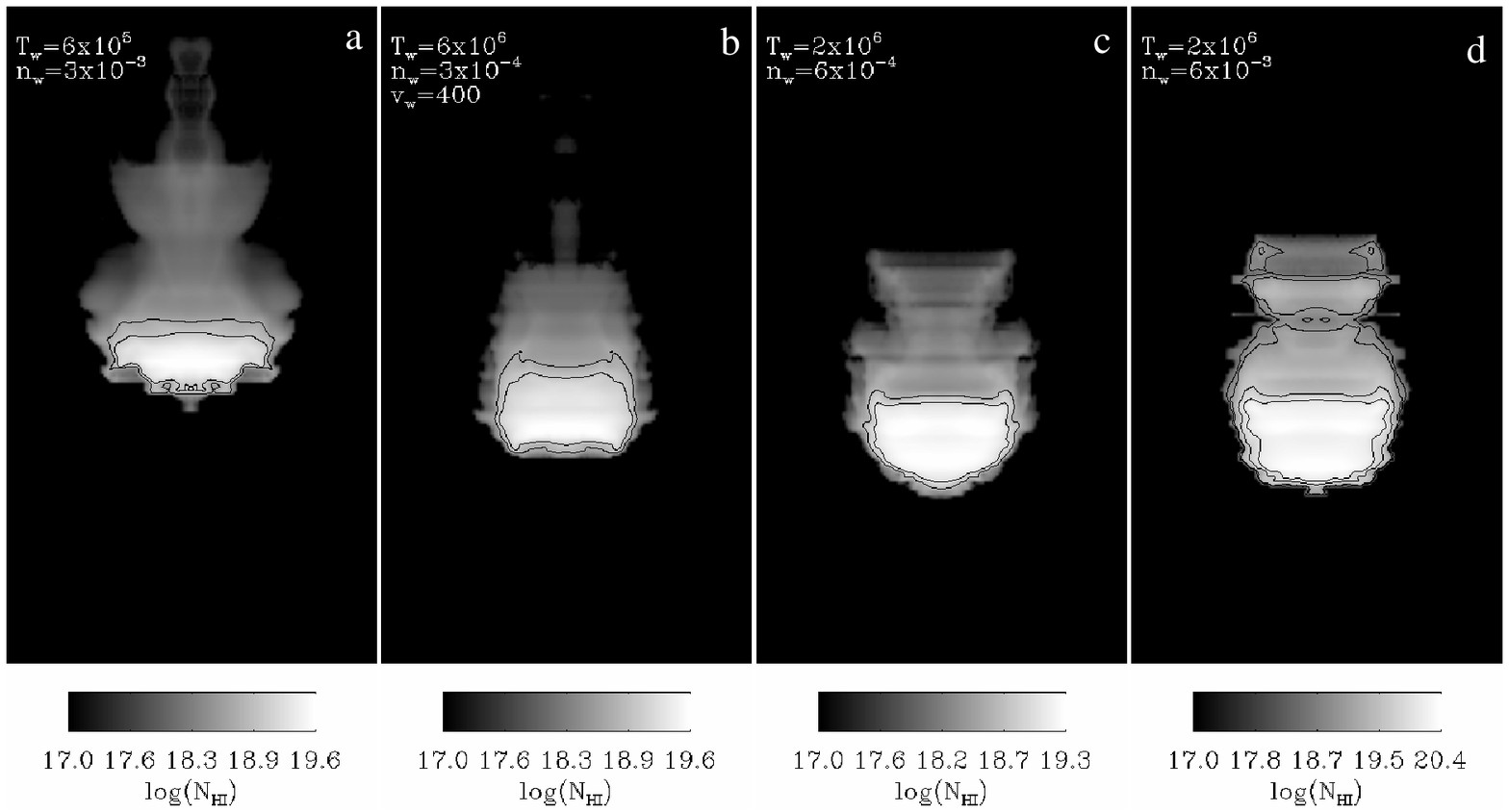}}
\caption{Column density maps ($N_{HI}$) for several of 
the pure gas HVCs after 3 Myrs of evolution. The temperature
and density of the external medium are indicated in each panel, and
in each case the wind velocity is 200 km/s except panel (b) which
is 400 km/s. The initial peak column density in HVCs (a)-(c) is
$2\times 10^{19}$ cm$^{-2}$ whereas model (d) was increased to 
$2\times 10^{20}$ cm$^{-2}$. The contours correspond to column densities 
$5, 10.0, 50.0, 100.0 \times 10^{18}$cm$^{-2}$.
} 
\label{fig3}
\end{figure*}

\section{Summary}

We have performed hydro-dynamical simulations of HVCs moving through a
diffuse medium in an attempt to explain observations of head-tail
structures and compression fronts. We find that these features are
reproduced in either pure gas clouds or gas embedded within a dark
matter halo, as long as the wind density is higher than $10^{-4}$
cm$^{-3}$.  This is supported by a simple application of Gunn \&
Gott's (1972) stripping criteria.

A variety of direct and indirect observations suggest that the density
of gas in the outer halo is $n_w(50{\rm kpc})\lsim 5\times 10^{-5}$
cm$^{-3}$.  Although observational constraints on the halo gas density
at 100--400 kpc are highly uncertain, our naive expectation is that it
should decrease at larger radii.  The ionised gas close to the
Galactic disk can reach much higher densities, $10^{-3}$ cm$^{-3}$,
therefore a Galactic fountain mechanism can reproduce these observations.
If the halo gas density increases between 50 and 300 kpc by a factor
of two then the observed head-tail structures could also be associated
with infalling extra-galactic dark matter dominated clouds.  

The lifetime of the neutral tails in our simulations is $\approx 10^9$
years after which time the column density drops below the
observational threshold. Since the observations suggest that
approximately 20\% of HVCs show evidence of tails or compression
fronts, a timing problem may exist since we must be witnessing the
destruction of a large fraction of the HVC population at the current
epoch. The Galactic fountain can circumvent this problem since 
clouds are continuously created and destroyed. The high fraction of clouds
with signs of interaction is expected since clouds are only visible as
they condense and fall back onto the disk at which point they are
susceptible to stripping.

Our simulations show that all HVCs should show tails of stripped gas
at lower column densities $\lsim 10^{18}$ cm$^{-2}$ which may be
detectable with higher resolution observations. One may also detect
$H_{\alpha}$ emission from the secondary shocks that heat the front of
the cloud.

\noindent{\bf Acknowledgments} \ We would like to thank 
Sergio Gelato, Mary Putman and Leo Blitz for useful 
discussions. VQ is a Marie Curie Fellow (grant HPMF-CT-1999-00052),
 BM is supported by the Royal Society. 
Simulations were carried out as part of the Virgo consortium on 
COSMOS an Origin 2000.

\end{document}